\documentclass[english,russian,12pt,twoside]{article}
\usepackage{amssymb,amsmath}
\usepackage[russian]{babel}
\usepackage[cp866]{inputenc}
\begin{document}
\begin{center} {\bf  ДИФФЕРЕНЦИАЛЬНАЯ АЛГЕБРА БИКВАТЕРНИОНОВ.\\ 4. ТВИСТОРЫ И ТВИСТОРНЫЕ
ПОЛЯ }
\end{center}
\vspace{3mm}
\begin{center}
{\bf Л.А. Алексеева }
\end{center}
\vspace{2mm}

\centerline{\textit{Институт математики и математического
моделирования КН МОН РК}}
 \centerline{\textit{Алматы, Казахстан, alexeeva@math.kz }}
\vspace{3mm}

Настоящая статья является продолжением статей [1-3], где
рассмотрены частные виды бикватернионных волновых уравнений,
эквивалентные системам уравнений Максвелла и Дирака и их
обобщениям, и построены их бикватернионные решения.

Здесь рассматривается бикватернионное волновое уравнение более
сложного вида, кото\-рое, если записать его в матричном
(тензорном) виде, относится к классу уравнений Янга-Милса [4].
Последние используются в теоретической физике для математического
описания элементарных частиц.

На основе дифференциальной алгебры бикватернионов и теории
обобщенных функций получены обобщенные решения биволнового
уравнения  при векторном пред\-став\-ле\-нии его структурного
коэффициента. Рассмотрены нестационарные, гармонические и
статические элементарные твисторы и  твисторные поля.

\bigskip

\textbf{4.1.  Твисторное биволновое уравнение и его решения.}
Рассмотрим линейные бикватернионные дифференциальные уравнения
вида:
\[
\nabla ^ \pm  {\bf B} + {\bf F} \circ {\bf B} = {\bf G}(\tau
,x),\,\,(\tau,x)\in \emph{M}, \eqno(1)
\]
где дифференциальные бикватернионные операторы $\nabla ^ \pm   =
\partial _\tau \pm i\nabla $ - взаимные биградиенты, действие которых определены алгеброй бикватернионов
[1]:
\[
\nabla ^ \pm  {\bf B}(\tau ,x) \buildrel \Delta \over = \left(
{\partial _\tau   \pm i\nabla } \right) \circ \left( {b(\tau ,x) +
B(\tau ,x)} \right) = \partial _\tau  b \mp i{\kern 1pt} {\rm
div}{\kern 1pt} B \pm i{\kern 1pt} {\rm grad}{\kern 1pt} {\kern
1pt} b + \partial _\tau  B \pm i{\kern 1pt} {\rm rot}{\kern 1pt} B
\]
соответственно верхнему (или нижнему) знаку,$\nabla$ -
градиент,$\emph{M}$ -- пространство Минковского, \emph{структурный
коэффициент} ${\bf F} = f + F$  - постоянный бикватернион,
\[{\bf F} \circ {\bf B}=(f+F)\circ(b+B)=fb-(F,B)+fB+bF+[F,B],
\] где $(.,.),[.,.]$ обозначают скалярное и векторное
произведение соответствующих векторов.

 Поскольку уравнения (1)
эквиваленты системам уравнений гиперболического типа и приводят к
решениям волновых уравнений, будем называть их \emph{биволновыми}
\emph{уравнениями общего вида}.

Частные случаи этого уравнения, когда ${\bf F} = 0$ и ${\bf F} =
f$- комплексное число, дают бикватернионное представление систем
уравнений Максвелла [2] и Дирака [3].

Здесь исследуем случай, когда ${\bf F} = F$ -- комплексный вектор.
Тогда (10) имеет вид (для верхнего знака):
\[
\left( {\partial _\tau   + i\nabla  + F} \right)\,{\bf B} = {\bf
G}.    \eqno(2)
\]
Рассмотрим взаимные операторы вида  $$ {\bf D}_F^ + = \nabla ^ + +
F,\quad {\bf D}_F^ - = \nabla ^ -   - F.
$$
Легко проверить, что их  суперпозиция  коммутативна и обладает
следующим полезным свойством:
\[
{\bf D}_F^ +  {\bf D}_F^ -   = {\bf D}_F^ -  {\bf D}_F^ +   =
\left( {\nabla ^ +   + F} \right) \circ \left( {\nabla ^ -   - F}
\right)\, =\Box  + (F,F) + 2i(F,\nabla ),  \eqno(3)
\]
где $\Box = \frac{{\partial ^2 }}{{\partial \tau ^2 }} - \Delta
$-- волновой оператор, $\Delta$ -- оператор Лапласа.  Используя
это свойство, из (2) получим
\[
{\bf D}_F^ -  {\bf D}_F^ +  {\bf B} = \left\{ { \Box+ (F,F) +
2i(F,\nabla )} \right\}{\bf B} = {\bf D}_F^ -  {\bf G} = {\bf Q}.
\]
Т.е. каждая компонента ${\bf B}$ удовлетворяет уравнению
\[
\Box u + (F,F)u + 2i(F,\nabla u) = q(\tau ,x)   \eqno(4)
\]
с соответствующей ${\bf Q}$ правой частью.

Далее построим  решения уравнения (1) для верхнего знака (+).
Решения (1)  для нижнего      знака (-) можно получить из них,
используя операцию комплексного сопряжения.
\smallskip

Т е о р е м а  4.1.1. \emph{Решение обобщенного биволнового
уравнения (2) можно представить в виде}:
\[
{\bf B} = {\bf D}_F^ -  \left( {\psi  * {\bf G}} \right) = {\bf
D}_F^ -  \psi  * {\bf G} + {\bf B}^0  = \psi  * {\bf D}_F^ -  {\bf
G} + {\bf B}^0, \eqno(5)
\]
\emph{ где $\psi (\tau ,x)$-- фундаментальное решение уравнения
(4) (при $q = \delta (\tau )\delta (x)$), а ${\bf B}^0 (\tau ,x)$
--  решение однородного уравнения  (2) (при ${\bf G} = 0$)}:
\[
{\bf B}^0  = \sum\limits_{\psi ^0 } {{\bf D}_F^ -  } \psi ^0  *
{\bf C}^0  = \sum\limits_{\psi ^0 } {\psi ^0  * {\bf D}_F^ -  }
{\bf C}^0  = \sum\limits_{\psi ^0 } {{\bf D}_F^ -  } \left( {\psi
^0  * {\bf C}^0 } \right) \eqno(6)
\]
 $\psi ^0 (\tau ,x)$--
\emph{решения однородного уравнения (4), $C^0 (\tau ,x)$ --
произвольные бикватернионы, допускающие такую свертку.}

Д о к а з а т е л ь с т в о: В силу линейности уравнения,
достаточно доказать      утверждение для каждого слагаемого в
формуле (5). Подставим первое  слагаемое  в уравнение (2) и,
используя  (3), получим
\[
\left( {\nabla ^ +   + F} \right)\left( {\nabla ^ -   - F}
\right)\left( {\psi  * {\bf G}} \right) = \left( {\Box\psi  +
2i(F,\nabla \psi ) + (F,F)\psi } \right) * {\bf G} = \delta (\tau
)\delta (x) * {\bf G} = {\bf G}.
\]
Для  каждого  слагаемого второй суммы имеем
\[
\left( {\nabla ^ +   + F} \right)\left( {\nabla ^ -   - F}
\right)\left( {\psi ^0  * {\bf C}^0 } \right) = \left\{  {\Box
\psi ^0  +2i(F,\nabla \psi ^0 ) + (F,F)\psi ^0 } \right\}
* {\bf C}^0  = 0.
\]
Здесь мы воспользовались известными свойствами сверток [5].
Очевидно, в силу линейности уравнения,  любое решение можно
представить в аналогичном виде.

Рассмотрим уравнение  (4), которое, если положить $m^2  = (F,F)$,
содержит оператор Клейна-Гордона-Фока  $\left( {\Box + m^2 }
\right)$ и дополнительное слагаемое ($ 2i(F,\nabla \psi )$ ) .
Интересно, что появление этого дополнительного члена значительно
упрощает вид его фундаментального решения,  в сравнении с
фундаментальным решением уравнения  Клейна-Гордона-Фока,
построенное Владимировым В.С. (см.в [6]).
\smallskip

Т е о р е м а 4.1.2.  \emph{Фундаментальные решения уравнения (4)
имеют вид}:
\[
\psi (\tau,x) = \frac{{e^{  i(F,x)} }}{{4\pi \left\| x
\right\|}}\left( {(1 - a)\delta (\tau  - \left\| x \right\|) +
a\delta (\tau  + \left\| x \right\|)} \right) + \psi ^0 ,\quad
\forall a \in \mathbb{Z},\eqno(7)
\]
\emph{где $\delta (\tau  \mp \left\| x \right\|)$ -- простой слой
на световом конусе  $\tau  =  \pm \left\| x \right\|$;  а
-произвольное комплексное число, $\psi ^0 (\tau ,x)$-- решение
однородного уравнения (при $ q = 0$). }

Д о к а з а т е л ь с т в о.  Для доказательства формулы теоремы
используем преобразование Фурье обобщенных функций.  Далее
переменные Фурье, соответствующие $(\tau ,x)$, обозначаем $\left(
{\omega ,\xi } \right)$ соответственно.

Уравнение для $\psi $ имеет вид:
\[
\Box\psi  + (F,F)\psi  +2i(F,\nabla \psi ) = \delta (\tau )\delta
(x), \eqno(8)\]
 а его преобразование Фурье
\[
\left( {\left\| \xi  \right\|^2  - \omega ^2  + 2(F,\xi ) + (F,F)}
\right)\bar \psi (\omega ,\xi ) = 1.\eqno (9)
\]
Откуда получим
   \[
\bar \psi (\omega ,\xi ) = \frac{1}{{\left( {\xi  + F,\xi  + F}
\right) - \omega ^2 }}. \eqno (10)
\]
Поскольку правая часть (10) имеет неинтегрируемые особенности, для
построения обратного преобразования Фурье следует выбрать
определенные регуляризации.

Для этого воспользуемся фундаментальным решением  уравнения
Даламбера:
\[
\Box \chi=\delta(\tau,x),
\]
которое имеет вид
\[\chi=
\frac{{1 - a}}{{4\pi \left\| x \right\|}}\delta (\tau  - \left\| x
\right\|) + \frac{a}{{4\pi \left\| x \right\|}}\delta (\tau  +
\left\| x \right\|),\quad \forall a\in \mathbb{Z}.
\]
Здесь $\delta (\tau  \pm \left\| x \right\|) $ -- простой слой на
световом конусе  $\tau  = \mp \left\| x \right\|$- сингулярная
обобщенная функция.

 Поскольку преобразование Фурье слагаемых
равно следующим регуляризациям :
\[
\texttt{F}\left[ {\frac{1}{{4\pi \left\| x \right\|}}\delta (\tau
\mp \left\| x \right\|)} \right] = \frac{1}{{\left\| \xi
\right\|^2 - \omega ^2  \pm i0}}, \eqno (11)
\]
используя свойства сдвига преобразования Фурье [5], из (10) и (11)
получим формулу теоремы (7). Теорема доказана.

Заметим, что  $\psi $ -- это сферическая волна, распространяющаяся
в  $R^3$ с единичной скоростью (если $\tau$ - время). При ${\rm
Re}\, F \ne 0$, реальная и мнимая часть плотности слоя на сфере
колеблются с изменением $x$, поэтому для уравнения (2) здесь
предложено такое название (автор не возражает, если найдется более
подходящее).  $\textrm{Im}\, F$  дает экспоненциальное затухание
или возрастание плотности в зависимости от направления $x$ по
отношению к $F$.\smallskip

\emph{Скалярные потенциалы}. Построим  решения однородного
уравнения (4):
\[
\Box  \psi ^0+ (F,F) \psi ^0 + 2i(F,\nabla \psi ^0) =0   \eqno(12)
\]
Его преобразование Фурье  имеет вид:
\[
\left( {\left( {\xi  + F,\xi  + F} \right) - \omega ^2 }
\right)\bar \psi ^0  = 0.
\]
 Следовательно $\bar \psi ^0  = \varphi (\omega ,\xi )\delta _S (\omega ,\xi ) $,
где   $\delta _S (\omega ,\xi )$
 -- простой слой на трехмерной поверхности $S$  в $R^4$:
\[
S = \left\{ {(\omega ,\xi ):\left( {\xi  + F,\xi  + F} \right) -
\omega ^2  = 0} \right\}, \eqno (13)
\]
 а $ \varphi (\omega ,\xi )$
 -произвольная интегрируемая на $S$ функция.

И формальное решение однородного уравнения имеет вид
поверхностного интеграла
\[
\psi ^0 (\tau ,x) = \int\limits_S {\varphi (\omega ,\xi )} \exp (
- i\omega \tau  - i(x,\xi ))dS(\omega ,\xi ),\quad\,\,\forall
\varphi (\omega ,\xi ) \in L_1 (S). \eqno (14)
\]
Рассмотрим, при каких $F$ такая поверхность существует и какой вид
она имеет.

Пусть  $F$ -- действительный вектор: $F=-E$. Тогда $S$ - это конус
в $R^4$ с вершиной в точке $\left( {\omega ,\xi } \right) = \left(
{0,E} \right)$. В этом случае (14) можно записать в виде:
\[
\psi ^0 (\tau ,x) = \int\limits_{R^3 } {\chi (\xi )} \exp \left( {
\pm i\left\| {\xi  - E} \right\|\tau  - i(x,\xi )}
\right)dV(\xi),\quad \forall \chi (\xi ) \in L_1 (R^3 ) \eqno
(15)\] $dV(\xi)=d\xi _1 d\xi _2 d\xi _3 .$

 Если $F$ - мнимый вектор: $F=-iH$. Тогда из (13) следует:
\[
\left( {\xi  - iH,\xi  - iH} \right) = \left\| \xi  \right\|^2  -
\left\| H \right\|^2  - 2i(H,\xi ) = \omega ^2
\]
Решением этого уравнения будет пересечение двух множеств,
задаваемых равенствами:
\[
S = \left\{ {(\omega ,\xi ):\left\| \xi  \right\|^2  - \left\| H
\right\|^2  = \omega ^2 ,\quad (H,\xi ) = 0} \right\}
\]
В этом случае решением однородного уравнения будет интеграл по
части плоскости, перпендикулярной вектору $H$, с выколотым кругом
радиуса $\left\| H \right\|$ с центром в точке $\xi=0$:
\[
\psi ^0 (\tau ,x) = \int\limits_{\scriptstyle {\xi \bot H}, \atop
\scriptstyle \left\| \xi  \right\| \ge \left\| H \right\|} {\chi
(\xi )} \exp \left( { \pm i\tau \sqrt {\left\| \xi  \right\|^2  -
\left\| H \right\|^2 }  - i(x,\xi )} \right)\,dS_\cap(\xi)  ,\quad
\forall \chi (\xi ) \in L_1 (S_\cap(\xi) ) \eqno (16)\]

    Если имеем комплексное $F=-E-iH$, тогда из (13) следует:
\[
\left( {\xi  - E - iH,\xi  - E - iH} \right) = \left\| {\xi  - E}
\right\|^2  - \left\| H \right\|^2  - 2i(\xi  - E,H) = \omega ^2
\]
В этом случае решением однородного уравнения будет интеграл по
части плоскости, проходящей через точку $\xi ^*  = E$  и
перпендикулярной вектору $H$, с выколотым кругом радиуса $\left\|
H \right\|$ с центром в $\xi ^*$:
\[
\psi ^0 (\tau ,x) = \int\limits_{\scriptstyle {(\xi-E) \bot H},
\atop \scriptstyle \left\| {\xi  - E} \right\| \ge \left\| H
\right\|} {\phi (\xi )} \exp \left( { \pm i\tau \sqrt {\left\|
{\xi  - E} \right\|^2  - \left\| H \right\|^2 }  - i(x,\xi )}
\right)\,dS_\cap(\xi) ,\quad \forall \phi (\xi ) \in L_1
(S_\cap(\xi)) \eqno(17)
\]
Здесь $$S_\cap(\xi)=\{\xi: {(\xi-E) \bot H} \cap\ { \| \xi  - E \|
\ge \| H \| }\} .$$  Выбор $\phi (\xi )$ позволяет строить широкий
класс решений твисторных  уравнений.

\bigskip

\textbf{4.2. Элементарный $\xi$-твистор. Нестационарные твисторные
поля.} Назовем решения однородного уравнения  (2)
\emph{твисторами}. Построим их бикватернионные представления.

Рассмотрим подынтегральные функции  в формуле решения (17)-
\[
\psi _\xi ^ \pm  (\tau ,x) = \exp \left( { \pm i\tau \sqrt
{\left\| {\xi  - E} \right\|^2  - \left\| H \right\|^2 }  -
i(x,\xi )} \right),\,\, \xi\in S_\cap(\xi)  \eqno (18)
\]
Они являются решением однородного уравнения (4) и представляют
собой две плоские гармонические волны, движущиеся в направлении
волнового вектора $\xi$  и противоположном направлении с фазовой
скоростью $c = \sqrt {\left\| {\xi  - E} \right\|^2  - \left\| H
\right\|^2 } /\left\| \xi  \right\|$; длина волн  $\lambda  = 2\pi
/\left\| \xi \right\|$.

При $\left\| {\xi  - E} \right\| > \left\| H \right\|$ их
$$\textrm{частота}\,\,\varpi = \sqrt {\left\| {\xi  - E} \right\|^2
- \left\| H \right\|^2 },\,\, \textrm{период } \,T = 2\pi /\sqrt
{\left\| {\xi - E} \right\|^2  - \left\| H \right\|^2 }.$$
Вычислим порождаемый ими элементарный $\xi$-твистор -
\[
{\bf \Psi }_\xi ^ \pm   = \frac{1}{{\sqrt 2 \left\| {\xi  - E}
\right\|}}{\bf D}_F^ -  \psi _\xi ^ \pm   =  \frac{{ \pm i\sqrt
{\left\| {\xi  - E} \right\|^2  - \left\| H \right\|^2 }  - ( \xi
- E) + iH }}{{\sqrt 2 \left\| {\xi  - E} \right\|}}\psi _\xi ^
\pm,\quad \xi\neq E \eqno(19)
\]
Его норма и псевдонорма равны
\[
\left\| {{\bf \Psi }_\xi ^ \pm  } \right\| = 1, \,\, \left\langle
{\left\langle {{\bf \Psi }_\xi ^ \pm  } \right\rangle }
\right\rangle  = i\frac{{\left\| H \right\|}}{{\left\| {\xi  - E}
\right\|}} \eqno (20)
\]
Бикватернион его энергии-импульса
$$\Xi ({\bf \Psi }_\xi ^ \pm ) = W({\bf \Psi }_\xi ^ \pm  ) +
iP({\bf \Psi }_\xi ^ \pm  ) = {\bf \Psi }_\xi ^ \pm   \circ \left(
{{\bf \Psi }_\xi ^ \pm  } \right)^* $$ равен:
 \[
 \Xi \left( {{\bf \Psi }_\xi ^ \pm  } \right) = 1 +
i\frac{{\left[ {e_{\xi  - E} ,H} \right] \pm e_{\xi  - E} \sqrt
{\left\| {\xi  - E} \right\|^2  - \left\| H \right\|^2 }
}}{{\left\| {\xi  - E} \right\|}},\quad   e_{\xi  - E}  =
\frac{{\xi  - E}}{{\left\| {\xi  - E} \right\|}}\eqno(21)
\]
Норма и псевдонорма $\Xi _\xi  $  равны
\[
\left\| {\Xi \left( {{\bf \Psi }_\xi ^ \pm  } \right)} \right\| =
\sqrt {1 + \frac{{\left\| {\xi  - E} \right\|^2  - \left\| H
\right\|^2 \cos ^2 \gamma }}{{\left\| {\xi  - E} \right\|^2 }}} ,
\quad \left\langle {\left\langle {\Xi \left( {{\bf \Psi }_\xi ^
\pm  } \right)} \right\rangle } \right\rangle  = \sqrt {1 -
\frac{{\left\| {\xi  - E} \right\|^2  - \left\| H \right\|^2 \cos
^2 \gamma }}{{\left\| {\xi  - E} \right\|^2 }}}
\]
где $\gamma $  угол между векторами $e_{\xi  - E} ,H$. Интересно,
что при  $\gamma=\pm\pi/2$
\[
\left\langle {\left\langle {\Xi \left( {{\bf \Psi }_\xi ^ \pm  }
\right)} \right\rangle } \right\rangle  = 0,\quad \left\| {\Xi
\left( {{\bf \Psi }_\xi ^ \pm  } \right)} \right\| = \sqrt 2,
\]
Наиболее простой вид ${\bf \Psi }_\xi ^ \pm  $  и $\Xi \left(
{{\bf \Psi }_\xi ^ \pm  } \right)$ имеют при  $H=0$  :
\[
{\bf \Psi }_\xi ^ \pm   = \frac{{ \pm i + e_{\xi  - E} }}{{\sqrt 2
}}\exp \left( { \pm i\tau \left\| {\xi  - E} \right\| - i(x,\xi )}
\right)\eqno (22)\]
\[ \left\| {{\bf \Psi }_\xi ^ \pm  } \right\| =
1,\quad \left\langle {\left\langle {{\bf \Psi }_\xi ^ \pm  }
\right\rangle } \right\rangle  = 0, \quad \Xi \left( {{\bf \Psi
}_\xi ^ \pm  } \right) = 1\pm i e_{\xi-E}.
\]
\smallskip

 Элементарные $\xi$-твисторы описывают  восьмимерные
плоские гармонические волны,  движущиеся  вдоль  вектора $\xi$, с
определенной амплитудой по каждой составляющей скаляр\-ной и
векторной действительной и мнимой части твистора.  Направление
движения опреде\-ля\-ется верхним либо нижним знаком твистора и
влияет на амплитуду волны.
\smallskip

\emph{Твисторные поля.} Используя ${\bf \Psi }_\xi ^ \pm $ можно
представить ${\bf B}^0 (\tau ,x)$ в виде суммы твисторов вида:
\[
\bf B^{\xi} (\tau ,x) =  {{\bf \Psi }^{+}_{\xi} \ast {\bf C}_1
(\tau ,x)+ \bf \Psi }^{-}_{\xi} \ast {\bf C}_2 (\tau ,x),
\eqno(23)
\]
которые описывают ${\xi}$-поляризованные твисторные поля.

Неполяризованные твисторные поля описываются бикватернионами вида
\[
\bf B^0 (\tau ,x) = \sum\limits_{C_0 ,\phi} {{\bf \Psi }^\phi
(\tau ,x)}  \ast {\bf C}^0 (\tau ,x), \eqno(24)
\]
\[
{\bf \Psi }^\phi  (\tau ,x) = \int\limits_{S_\cap} {\phi (\xi
){\bf \Psi }_\xi } (\tau ,x)dS_\cap(\xi ),\quad \forall \phi  \in
L_1 (S_\cap) .
\]
Скалярно-векторные поля $\textbf{C}_j(\tau ,x)$ произвольные,
допускающие такие свертки, в том числе могут быть из класса
сингулярных обобщенных функций.

Заметим, что выше мы построили решения твисторного уравнения, допускающие обобщенное преобразование Фурье.
Однако, существуют его решения. не принадлежащие обобщенным функциям медленного роста.
\smallskip

\emph{Экпоненциально затехающие и возрастающие по времени твисторы.} При $\left\| {\xi  - E} \right\| \leq \left\| H
\right\|$ скалярный потенциал (18)  преобразуется к виду:
\[
\alpha _\xi ^ \pm  (\tau ,x) = \exp \left( { \pm \tau \sqrt {
\left\| H \right\|^2 -\left\| {\xi  - E} \right\|^2  }  - i(x,\xi
)} \right),\,\, \xi-E \bot H,  \eqno (25)
\]
и также является решением однородного уравнения (12). Это две
стоячие гармонические волны с экспоненциально затухающей и
возрастающей по времени амплитудой. При $\left\| {\xi - E}
\right\| = \left\| H \right\|$ амплитуда не зависит от времени и
они совпадают.

Вычислим порождаемый (25) элементарный твистор:
\[
{\bf \Upsilon }_\xi ^ \pm (\tau,x)  = \frac{{\bf D}_F^ - \alpha
_\xi ^ \pm }{{\left\| {\xi  - E} \right\|\sqrt 2 }}  = \frac{{ \pm
\sqrt { \left\| H \right\|^2-\left\| {\xi  - E} \right\|^2   }  -
( \xi - E) + iH }}{{\left\| {\xi  - E} \right\|\sqrt 2 }}\alpha
_\xi ^ \pm (\tau,x),\quad \xi-E\bot H \eqno(26)
\]
Его норма и псевдонорма равны
\[
\left\| {{\bf \Upsilon }_\xi ^ \pm  } \right\| =\frac{{\left\| H
\right\|}}{{\left\| {\xi  - E} \right\|}}, \,\, \left\langle
{\left\langle {{\bf \Upsilon }_\xi ^ \pm  } \right\rangle }
\right\rangle  = i \eqno (27)\] Бикватернион его энергии-импульса
равен:
 \[
 \Xi \left( {{\bf \Psi }_\xi ^ \pm  } \right) = 1 +
\frac{{i\left[ {e_{\xi  - E} ,H} \right] \pm e_{\xi  - E} \sqrt {
\left\| H \right\|^2-\left\| {\xi  - E} \right\|^2   } }}{{\left\|
{\xi  - E} \right\|}} \eqno(28)
\]

Аналогично (23),(24) можно строить эспоненциально возрастающие или
затухающие  по времени твисторные поля, из  поляризованных вида
\[
\bf B^{\xi} (\tau ,x) ={\bf \Upsilon }^{\pm}_{\xi} \ast {\bf C}
(\tau ,x),
\]
или неполяризованные вида
\[
\bf B^0 (\tau ,x) = \sum\limits_{C ,\phi} {{\bf \Upsilon }^\phi
(\tau ,x)}  \ast {\bf C} (\tau ,x),
\]
\[
{\bf \Upsilon }^\phi  (\tau ,x) = \int\limits_{S_\cap} {\phi (\xi
){\bf \Upsilon }_\xi } (\tau ,x)dS_\cap(\xi ),\quad \forall \phi
\in L_1 (\xi\bot  H) .\]

 Если $\xi=E$, из скалярного
потенциала (18) получим элементарный $H$-твистор:
\[
 {\bf \Psi }_H^ \pm   = \frac{1}{\sqrt{2}{\left\| H \right\| }}{\bf
D}_F^ -  \exp \left(  \mp\tau \| H\| - i(E,x ) \right)  =
\frac{(\mp 1  + i e_H)}{\sqrt{2}}\exp \left(  \mp\tau \| H\| -
i(E,x ) \right),\]
\[ \left\| {{\bf \Psi }_H^ \pm  } \right\| =
1 ,\quad \left\langle {\left\langle {{\bf \Psi }_H^ \pm  }
\right\rangle } \right\rangle = 0 ,\quad {\bf \Xi }\left( {{\bf
\Psi }_H^ \pm  } \right) = 0.
\]

\bigskip

\textbf{4.3. Стационарные  решения твисторных уравнения.}
Рассмотрим также важный для приложений  класс решений уравнения
(11) вида $${\bf B} = {\bf B}(x)e^{ - i\omega \tau }, $$ которые
описывают гармонические колебания с частотой   $\omega$.
Предполагается, что правая часть (11) имеет ту же структуру: ${\bf
G} = i{\bf G}(x)e^{ - i\omega \tau } $. Тогда для комплексных
амплитуд получим уравнение, которое приводится к виду:
\[
\left( {\nabla _\omega ^{^ \pm  }  + F} \right)\,{\bf B} = {\bf
G},
\eqno (24)
\]
где введены \emph{взаимные $\omega$-градиенты} $\nabla _\omega ^
\pm   = \omega \pm \nabla $. Решения соответствующего однородного
уравнения назовем $\omega$-\emph{твисторами. }

Также прямым вычислением доказывается следующее свойство:
\[
\left( {\nabla _\omega ^ \pm   \pm F} \right) \circ \,\left(
{\nabla _\omega ^ \mp   \mp F} \right) = \Delta  \pm 2(F,\nabla )
+ \omega ^2  + (F,F).
\]
Используя его, получим
\[
\left( {\nabla _\omega ^ \mp   - F} \right)\left( {\nabla _\omega
^ \pm   + F} \right)\,{\bf B} = \left[ {\Delta  \pm 2(F,\nabla ) +
\omega ^2  + (F,F)} \right]\,{\bf B} = \left( {\nabla _\omega ^
\mp   - F} \right){\bf G} = {\bf Q}
\]
Т.е. каждая компонента является решением уравнения
\[
\left( {\Delta  \pm 2(F,\nabla ) + \omega ^2  + (F,F)} \right)\psi
_\omega   = q(x) \eqno (25)
\]
(в соответствии верхнему или нижнему знакам) с соответствующей
${\bf Q}$ правой частью.

 Построим фундаментальное решение этого
уравнения  с помощью  обобщенного преобразова\-ния Фурье по $x$.
Его Фурье-трансформанта имеет вид:
\[
\bar \psi _\omega   = \frac{1}{{\omega ^2  - \left\| \xi
\right\|^2  - 2i(F,\xi ) + (F,F)}} = \frac{1}{{\omega ^2  - (\xi +
iF,\xi  + iF)}}.
\]
Используя фундаментальное решение уравнения Гельмгольца [5]:
\[
\Delta \chi  + \omega ^2 \chi  = \delta (x),
\]
\[
\chi  =  - \frac{1}{{4\pi \left\| x \right\|}}\left( {ae^{i\omega
\left\| x \right\|}  + (1 - a)e^{ - i\omega \left\| x \right\|} }
\right),\quad \bar \chi  =  - \frac{a}{{\left\| \xi  \right\|^2  -
\omega ^2  + i0}} - \frac{{1 - a}}{{\left\| \xi  \right\|^2  -
\omega ^2  - i0}},
\]
и свойства преобразования Фурье, получим
\[
\psi _\omega  (x) = \frac{{e^{ - (F,x)} }}{{4\pi \left\| x
\right\|}}(ae^{i\omega \left\| x \right\|}  + (1 - a)e^{ - i\omega
\left\| x \right\|} ). \eqno (26)
\]
На его основе аналогично, как в нестационарном случае,
доказывается следующая теорема.\vspace{2mm}

Т е о р е м а  4.3.1. \emph{Решения уравнения }(24) \emph{ можно
представить в виде суммы бикватернионов:}
\[
{\bf B} = \left( {\nabla _\omega ^ \mp   - F} \right)\left( {\psi
_\omega   * } \right.\left. {\bf G} \right) + {\bf T}_\omega \eqno
(27)
\]
\emph{где  $\psi ^\omega  (x)$  имеет вид} (26), \emph{а  ${\bf
T}_\omega $- решение однородного  уравнения}
\[
\left( {\nabla _\omega ^{^ \pm  }  + F} \right)\,{\bf T}^\omega
= 0 ,     \eqno (28)
\]
\emph{которое можно представить в виде:}
\[
{\bf T}^\omega   = \left( {\nabla _\omega ^ \mp   - F}
\right)\left( {\psi _\omega ^0  * {\bf C}(x)} \right),\quad\eqno
(29)
\]
$\forall {\bf C}(x)$, \emph{допускающее эту свертку,  $\psi
_\omega ^0 $ - решение однородного уравнения:}
\[
\left( {\Delta  \pm 2(F,\nabla ) + \omega ^2  + (F,F)} \right)\psi
_\omega ^0  = 0,   \eqno (30)
\]
\emph{либо в виде суммы решений вида }(29).

Построим   $\omega$-твисторы   ${\bf T}^\omega  $ (для верхнего
знака).  Используя  преобразования Фурье по $x$ из (30) получим:
\[
\left( {\omega ^2  - (\xi  + iF,\xi  + iF)} \right)\bar \psi
_\omega   = 0.   \eqno (31)
\]
Следовательно $\bar \psi _\omega ^0  = \varphi (\xi )\delta
_{S^\omega  } (\xi )$, где $\varphi (\xi )$-произвольная локально
интегрируемая функция, а $\delta _{S^\omega  } (\xi )$ -- простой
слой на поверхности $S^\omega  $ в $R^3 $:
\[
S^\omega   = \left\{ {\xi :\left( {\xi  + iF,\xi  + iF} \right) =
\omega ^2 } \right\}.
\]
И формальное решение однородного уравнения имеет вид интеграла
\[
\psi ^0 (x) = \int\limits_{S^\omega  } {\varphi (\xi )} \exp ( -
i(x,\xi ))dS^\omega  (\xi ),    \quad \forall \varphi (\xi ) \in
L_1 (S^\omega  ) .  \eqno (32)
\]
Определим ${S^\omega  } $.

Если ${\mathop{\rm Im}\nolimits} \,F = 0$, т.е. $F$ -действительный
вектор: $F=E$, тогда
\[
S^\omega   = \left\{ {\xi :\left\| \xi  \right\|^2  = \omega ^2  +
\left\| E \right\|^2  \cap \left( {\xi ,E} \right) = 0} \right\}.
\]
В $R^3$ это окружность -- пересечение сферы радиуса  $r^*  = \sqrt
{\omega ^2  + \left\| E \right\|^2 } $  с плоскостью $\left( {\xi
,E} \right) = 0$ . В этом случае в формуле (32)  интеграл
контурный по этой окружности:
\[
\psi ^0 (x) = \int\limits_{\left\{ {e \bot E} \right\}} {\beta
(e)} \exp \left( { - i(x,e)\sqrt {\omega ^2  + \left\| E
\right\|^2 } } \right)dl(e),\quad \left\| e \right\| = 1,\quad
\eqno (33)
\]
$\forall \beta (e) \in L_1 \left\{ {e \in R^3 :\left\| e \right\|
= 1,\;e \bot E} \right\}$.

Если $F$ - мнимый вектор: $F=iH$, тогда из (31) следует, что
${S^\omega  } $ -- это сфера: $\left\| {\xi  - H} \right\| =
\left| \omega  \right|$  с центром $\xi ^*  = H$, и интеграл (32)
приводится к виду:
\[
\psi ^0 (x) = \exp ( - i(x,H))\int\limits_{\left\| e \right\| = 1}
{\alpha (e)} \exp ( - i\omega (x,e))dS(e),\quad \forall \alpha (e)
\in L_1 \{ e:\left\| e \right\| = 1\}. \eqno (34)
\]

Если имеем комплексное $F = E + iH$, тогда
\[
S^\omega   = \left\{ {\xi :\left\| {\xi  - H} \right\|^2  = \omega
^2  + \left\| E \right\|^2  \cap (E,\xi  - H) = 0} \right\}
\]
В $R^3$ это окружность -- пересечение сферы радиуса  $r^*  = \sqrt
{\omega ^2  + \left\| E \right\|^2 } $  с центром в точке $\xi  =
\pm H$ с плоскостью $\left( {\xi ,E} \right) = (H,E)$ . В этом
случае в формуле (33)  интеграл контурный по этой окружности:
\[
\psi ^0 (x) = e^{ - i(x,H)} \int\limits_{\left\{ {e \bot E}
\right\} \cup \{ \left\| e \right\| = 1\} } {\gamma (e)} \exp
\left( { - i(x,e)\sqrt {\omega ^2  + \left\| E \right\|^2 } }
\right)dl(e), \eqno (35)
\]
$e_\xi   = \xi /\left\| \xi  \right\|,\quad \forall \gamma (e )
\in L_1 \{ e:\left\| e \right\| = 1,e \bot E\} $.  Выбор $\gamma
(e )$  позволяет строить  широкий  класс
$\omega$-твисторов.\vspace{2mm}

    \emph{Элементарные $\omega$-твисторы}. Рассмотрим функцию
\[
\psi ^\omega  (x) = \exp \left( { - i(x,H + e_E \sqrt {\omega ^2
+ \left\| E \right\|^2 } )} \right)\quad ,\quad e_E  \bot E,\quad
\left\| {e_E } \right\| = 1, \eqno (36)
\]
которая является решением однородного уравнения (30) и
представляет собой комплексную амплитуду плоских гармонических
волн (с учетом $e^{ \pm i\omega \tau } $),  движущихся в
направлении волнового вектора $K_F  = e_E \sqrt {\omega ^2  +
\left\| E \right\|^2 }  + H$  или  противоположном (для
$e^{i\omega \tau } $) со скоростью $v = \omega /\left\| {K_F }
\right\|$; длина волн $\lambda  = 2\pi /\left\| {K_F } \right\|$,
частота $\varpi  = \omega $, период колебаний $T = 2\pi /\omega $.

При $H=0$  скорость $v = \omega /\sqrt {\omega ^2  + \left\| E
\right\|^2 }  \le 1$, не зависит от направления волны, с ростом
$\left\| E \right\|$ длина волны $\lambda  = 2\pi /\sqrt {\omega
^2  + \left\| E \right\|^2 } $  уменьшается; при $ \omega  \to
\infty \;\;v \to 1,\;\;\lambda  \to 0.$

При $E=0$
 $v = \frac{\omega }{\left\| H \right\|},\quad
 \lambda  =
\frac{{2\pi }}{\left\| H \right\|} $; при $ \left\| H \right\| \to
\infty$       $ v \to 0,\,\lambda \to 0 $  . \vspace{3mm}

Рассмотрим элементарный $\omega$-твистор -
\[
{\bf \Psi }_ + ^\omega   = \frac{1}{{\sqrt 2 \sqrt {\omega ^2  +
\left\| E \right\|^2 } }}\left( {\nabla _\omega ^ -   - F}
\right)\psi ^\omega   = \frac{{\omega  - E + ie_E \sqrt {\omega ^2
+ \left\| E \right\|^2 } }}{{\sqrt 2 \sqrt {\omega ^2  + \left\| E
\right\|^2 } }}\psi ^\omega,\quad e_E\bot E \eqno (37)
\]
Его амплитуда не зависит от $H$, норма и псевдонорма равны:
\[
\left\| {{\bf \Psi }_ + ^\omega  } \right\| = 1 ,        \quad
\left\langle {\left\langle {{\bf \Psi }_ + ^\omega  }
\right\rangle } \right\rangle  =  \frac{{i\left\| E
\right\|}}{{\sqrt {\omega ^2  + \left\| E \right\|^2 } }}. \eqno
(38)
\]
Бикватернион его энергии-импульса равен:
\[
\displaylines{ \Xi _\omega   = W_\omega   + iP_\omega   = {\bf
\Psi }_\omega ^ +   \circ \left( {{\bf \Psi }_\omega ^ +  }
\right)^*  = 1 + i\frac{{\omega e + [e,E]}}{{\sqrt {\omega ^2  +
\left\| E \right\|^2 } }} \cr \left\| {\Xi _\omega  } \right\| =
\sqrt 2 ,\quad \left\langle {\left\langle {\Xi _\omega  }
\right\rangle } \right\rangle  = 0 \cr}
\]
При  $E=0$   имеем  \[ {\bf \Psi }_ + ^\omega   = \left( {1 + ie}
\right)\exp ( - i(x,H + e\omega )) ,   \quad \left\| {{\bf \Psi }_
+ ^\omega  } \right\| = 1,  \quad \left\langle {\left\langle {{\bf
\Psi }_ + ^\omega  } \right\rangle } \right\rangle  = 0.
\]

Используя ${\bf \Psi }_\omega ^ + $ , можно представить ${\bf
T}_\omega  (x)$  в виде суммы твисторов типа:
\[
{\bf T}_\omega  (x) = \sum\limits_C {{\bf \Psi }_\omega  }  * {\bf
C}(x)  \eqno (39)
\]
\[
{\bf T}_\omega  (x) = \sum\limits_{C,\phi } {{\bf \Psi }_\omega
^\phi  }  * {\bf C}(x),\quad {\bf \Psi }_\omega ^\phi   =
\int\limits_{S_\omega  } {\phi (e){\bf \Psi }^\omega_ +  }
(e,x)dl(e),\quad \forall \phi  \in L_1 \left\{ {e \bot E:\left\| e
\right\| = 1} \right\}.     \eqno (40)
\]
Скалярно-векторные поля  ${\bf C}(x)$ тоже произвольные,
допускающие такие свертки, в том числе могут быть  из класса
сингулярных обобщенных функций.

Твисторы вида  (39)  описывают ориентированные колебания и волны
скалярно-векторных полей, а  (40) - неориентированные поля.
\bigskip

\textbf{4.4. Статические  решения твисторных уравнений.
Статические твисторы}

\emph{Статический твистор} - это решение уравнения  (1),  не
зависящее от времени (${\bf G} = {\bf G}(x)$). Его легко получить
из ${\bf T}^\omega  $ при $\omega  = 0$, т.е. в наших обозначениях
это ${\bf T}^0  $. Выпишем его отдельно.

Преобразования Фурье по $x$ (30) в этом случае имеет вид :
\[
\left( {\xi  + iF,\xi  + iF} \right)\bar \psi _\omega   = 0.
\]
И формальное решение однородного уравнения имеет вид интеграла
\[
\psi ^0 (x) = \int\limits_{S^\omega  } {\varphi (\xi )} \exp ( -
i(x,\xi ))dS^0 (\xi ),\quad \forall \varphi (\xi ) \in L_1
(S^\omega  ) . \eqno (41)
\]
Если $\textrm{Im} \,F=0$, т.е. действительный вектор: $F=E$, тогда
\[
S^0  = \left\{ {\xi :\left\| \xi  \right\|^2  = \left\| E
\right\|^2  \cap \left( {\xi ,E} \right) = 0} \right\}.
\]
В $R^3$ это окружность радиуса  $r^*  = \left\| E \right\|$--
пересечение сферы радиуса  $\left\| E \right\|$ с плоскостью
$\left( {\xi ,E} \right) = 0$ . В этом случае в формуле (41)
интеграл контурный по этой окружности:
\[
\psi ^0 (x) = \int\limits_{\left\{ {e \bot E} \right\}} {\beta
(e)} \exp \left( { - i\left\| E \right\|(x,e)} \right)dl(e),\quad
\left\| e \right\| = 1,\quad \eqno (42)
\]                  \
$\forall \beta (e) \in L_1 \left\{ {e \in R^3 :\left\| e \right\|
= 1,\;e \bot E} \right\}$.

Если $F$ - мнимый вектор: $E=iH$, тогда $S^0$: $\left\| {\xi  - H}
\right\| = 0 \Rightarrow \xi  = H$,
\[
\bar \psi  = a\delta (\xi  - H)\quad  \Rightarrow \quad \psi ^0
(x) = a{\mathop{\rm e}\nolimits} ^{ - i(x,H)}. \eqno (43)
\]

    Если имеем комплексное $F=E+iH$, тогда
\[
S^0  = \left\{ {\xi :\left\| {\xi  - H} \right\| = \left\| E
\right\| \cap (E,\xi  - H) = 0} \right\}.
\]
В $R^3$ это окружность радиуса  $r^*  = \left\| E \right\|$--
пересечение сферы радиуса  $r^*  $  с центром в точке $\xi  =  \pm
H$ с плоскостью $\left( {\xi ,E} \right) = (H,E)$ . В этом случае
в формуле (42)  интеграл контурный по этой окружности:
\[
\psi ^0 (x) = e^{ - i(x,H)} \int\limits_{\left\{ {e \bot E}
\right\} \cup \{ \left\| e \right\| = 1\} } {\gamma (e)} \exp
\left( { - i\left\| E \right\|(x,e)} \right)dl(e),\quad \forall
\gamma (e) \in L_1 \{ e:\left\| e \right\| = 1,e \bot E\}.
\eqno(44)
\]

\emph{Элементарные статические твисторы}. Соответствующий
потенциал в этом случае имеет вид синусоиды вдоль направления
$e=E+H$:
\[
\psi ^0 (x) = \exp \left( { - i(x,E + H)} \right),
\]
период которой равен $\frac{{2\pi }}{{\left\| {E + H} \right\|}}$.
Порождаемый им  элементарный статический твистор -
\[
{\bf \Psi }_ + ^0  = \frac{1}{{\sqrt 2 \left\| E \right\|}}\left(
{\nabla   - F} \right)\psi ^0  = \frac{{ - 1 + i}}{{\sqrt 2 }}\psi
^0 e_E ,\quad \left\| {{\bf \Psi }_ + ^\omega } \right\| =
1,\;\;\left\langle {\left\langle {{\bf \Psi }_ + ^\omega  }
\right\rangle } \right\rangle  = -1. \eqno (45)
\]
Бикватернион его энергии-импульса равен:
\[
\Xi ^0  = 1 + i[e,e_E ],\quad \left\| {\Xi ^0 } \right\| = \sqrt 2
,\quad \left\langle {\left\langle {\Xi ^0 } \right\rangle }
\right\rangle   = 0
\]
Статический твистор можно представить в виде
 \[
{\bf T}^0 (x) = \sum\limits_{C (x)} {{\bf \Psi }^0 } (e_E ,x) *
{\bf C}(x)
\]
 либо
\[
\displaylines{ {\bf T}^0 (x) = \sum\limits_{C ,\phi } {{\bf \Psi
}_\phi ^0 }  * {\bf C}(x),\quad  \cr {\bf \Psi }_\phi ^0  =
\int\limits_{\scriptstyle \left\| e \right\| = 1, \hfill \atop
\scriptstyle e \bot E \hfill} {\phi (e){\bf \Psi }^0 }
(e,x)dl(e),\quad \forall \phi  \in L_1 \left\{ {e:\left\| e
\right\| = 1,e \bot E} \right\} .\cr}
\]
Здесь $\textbf{C}(x)$  - произвольные регулярные или сингулярные
бикватернионы, допускающие такие свертки.
\bigskip

 \textbf{Заключение.}  Здесь показано, что для твисторов существуют порождающие их
скаляр\-ные потенциалы, которые удовлетворяют уравнению (30).
Решения этого уравнения пред\-став\-ляют\-ся в виде поверхностных
и контурных интегралов от элементарных потенциалов, выражаемых
через экспоненциальные функции, и содержат также достаточно
произ\-воль\-ные подын\-тег\-раль\-ные функции типа $\alpha(e),
\beta(e)$ . От этих представлений нетрудно перейти к представлению
твисторов с использованием функций Бесселя и сферических гармоник,
если выбирать их соответственно интегральным разложениям этих
специальных функ\-ций. В этом случае можно получить счетное число
еще более элементарных твисторов, которые можно использовать в
теории элементарных частиц.

Однородное уравнение (1) также можно рассматривать как частный
случай уравнения трансформации масс-зарядов, электрических и
гравимагнитных токов во внешнем электро-гравимагнитном (ЭГМ) поле
[7-9], напряженность которого описывается комплексным вектором
$F$. В этом случае полученным решениям можно придать физическую
трактовку, согласно введенным там бикватернионам заряда-тока,
мощности-силы ЭГМ-полей. Но наиболее  полным такое представление
будет при решении бикватернионного уравнения общего вида (1),
когда $\textbf{F}$ содержит скалярную и векторную часть
($\textbf{F}=f+F$). С решением этой задачи автор познакомит
читателя в последней статье этого цикла.

\bigskip

\emph{Ключевые слова}: бикватернион, биволновое уравнение,
векторный коэффициент, обоб\-щен\-ное решение, твистор.

\bigskip
\emph{Key words}: biquaternion, biwave equation, vector factor,
generalized solution, twistor.

 \vspace{5mm} MSQ 46S10, 53C80
\vspace{10mm}

 \centerline {СПИСОК ИСПОЛЬЗОВАННЫХ
ИСТОЧНИКОВ}

 1. Алексеева Л.А. Дифференциальная алгебра бикватернионов. 1. Преобразования
Лоренца// Математический журнал.-2010.-Т.10.-№.1.-С.33-41 10

2.  Алексеева Л.А. Дифференциальная алгебра бикватернионов. 2.
Обобщенные решения биволновых уравнений//   Математический
журнал.- 2010. -Т.10. -№.4. -С.5-13

3.  Алексеева Л.А.Дифференциальная алгебра бикватернионов. 3.
Уравнение Дирака и его обобщенные решения//Математический
журнал.-2011. -Т.11.-№1.-С.33-41

4. Yang C.N.,Mills R. Conservation of Isotropic Spin and Isotropic
Gauge Invariance//Physical review.-1954-96(1)-P.191-195

5.  Владимиров В.С. Обобщенные функции в математической физике.
-М. -1976.-512c.

6.  Полянин А.Д. Справочник по линейным уравнениям математической
физики - М.: ФИЗМАТЛИТ, -2001. - 576 с.

7.  Алексеева Л.А. Уравнения взаимодействия А-полей и законы
Ньютона  //Известия НАН РК. Серия физико-математическая. -2004.
-№3. -С.45-53.

8. Алексеева Л.А. Преобразования Лоренца для одной модели
электро-гравимагнитного поля.  Законы сохранения// Математический
журнал.- 2007.-Т.7.-№4.-С.12-24.

9.Алексеева Л.А. Полевые аналоги законов Ньютона для одной модели
электро-грави\-маг\-нит\-ного поля//Гиперкомплексные числа в
геометрии и физике. -2009. -Т6. -№ 1. -С.122-134
\bigskip

\emph{Алексеева Людмила Алексеевна}, проф. д.ф.-м.н.

ИМММ КН МОН РК,

050010 Казахстан, Алма-Ата 100, ул. Пушкина 125,

 E-mail: alexeeva@math.kz

 Моб.тел. +7-7773381814

\newpage

\textbf{Алексеева Л.А. Дифференциальная алгебра бикватернионов.
4.Твисторы и твисторные поля}// Математический
журнал.-2013.-Т.13.-N....-С.\bigskip

На основе дифференциальной алгебры бикватернионов и теории
обобщенных функций рассмотрено бикватернионное волновое
(\emph{биволновое}) уравнение общего вида при векторном
пред\-став\-ле\-нии его структурного коэффициента. Если записать
это уравнение в матричном (тензорном) виде, оно относится к классу
уравнений Янга-Милса, которые используются в теоретической физике
для математического описания элементарных частиц. Построены его
обобщенные решения, описывающие нестационарные, гармонические и
статические элементарные твисторы и  твисторные поля.

\bigskip
\textbf{Alexeyeva L.A. Differential algebra of biquaternions. 4.
Twistor and twistors fields}//Mathe\-mati\-cal journal.-2013.-V.
13.-No...-P.\bigskip

On base of the differential biquaternions algebra  and theories of
generalized function the  biquaternionic  wave (\emph{biwave})
equation of  general type is considered under vector
representation of its structural factor. If to write this equation
in matrix (tensor) form, it pertains to class of the Young-Mils
equations , which are used in theoretical physics  for
mathematical description of the elementary particles. Its
generalized decisions, describing nonstationary, harmonic and
static elementary twistors and twistor fields, are built.

\end{document}